\documentclass[aps,prb,titlepage,twocolumn,floatfix,showpacs]{revtex4}

\usepackage{amssymb}
\usepackage{amsfonts}
\usepackage{amsmath,amsthm}
\usepackage{graphicx}
\usepackage{bm}
\usepackage{float}
\usepackage{amsmath}
\usepackage[T1]{fontenc}
\usepackage[latin1]{inputenc}
\usepackage{times}
\usepackage[normalem]{ulem}
\usepackage{setspace}
\usepackage{subfigure}
  \makeatletter
  \def\@dotsep{4.5}
  \makeatother
\newlength{\myVSpace}
\newcommand{\refon}[1]{Ref.~[\onlinecite{#1}]}

\newcommand{\be}{\begin{equation}}
\newcommand{\ee}{\end{equation}}
\newcommand{\ba}{\begin{eqnarray}}
\newcommand{\ea}{\end{eqnarray}}

\begin{document}

\title{
Effect of confinement potential geometry on entanglement in quantum dot-based nanostructures  
}
\author{S. Abdullah}
\email{sma503@york.ac.uk}

\author{J. P. Coe}
\email{jpc503@york.ac.uk}

\author{I. D'Amico}
\email{ida500@york.ac.uk}

\affiliation{
Department of Physics, University of York, York YO10 5DD, United
Kingdom
}

\begin{abstract}
We calculate the  spatial entanglement between two electrons trapped in a nanostructure for a broad class of confinement potentials, including single and double quantum dots, and core-shell quantum dot structures.
By using a parametrized confinement potential, we are able to switch from one structure to the others with continuity and
to analyze how  the entanglement is influenced by the changes in the  confinement geometry.  
We calculate the  many-body wave function  by `exact' diagonalization of the time independent Schr\"odinger equation. 
We discuss the relationship between the entanglement and specific cuts of the wave function, and show that the wave function at a single highly symmetric point could be a good indicator for the entanglement content of the system. We analyze the counterintuitive relationship between spatial entanglement and Coulomb interaction, which connects maxima (minima) of the first to minima (maxima) of the latter. We introduce a potential quantum phase transition which relates quantum states characterized by different spatial topology. 
Finally we show that by varying shape, range and strength of the confinement potential, it is possible to induce strong and rapid variations of the entanglement between the two electrons.  This property may be used to tailor nanostructures according to the level of entanglement required by a specific application. 
\end{abstract}
\maketitle

\section{Introduction}
Entanglement is a quantum property which provides the possibility for quantum information/computation to overcome some of the limitations of traditional devices. For this reason entanglement is now considered a physical resource.  Recently, semiconductor quantum dots (QDs) have been proposed as promising hardware to perform quantum information/computation within solid state.\cite{QDSchemes,LdV}  Their potential advantages include the existence of an industrial base for semiconductor processing and flexibility in driving the computational degrees of freedom by applied electro-magnetic fields and purposely designed trains of laser pulses.\cite{laser}  The system parameters may be tuned, making it possible to tailor the properties of semiconductor nanostructures.\cite{tune,LdV}  The rapid technological advances seem to promise sophisticated engineering of QD-based structures, with the potential for the production of scalable and coupled QD systems.\cite{double_gate_dots,experimental_papers}  
A crucial requirement is then the possibility of generating and manipulating entanglement within these structures.
  
In QD systems entanglement could be controlled by externally applied  electro-magnetic fields,\cite{e-m_entangl} or by varying nanostructure parameters.
Recently the effect of the interdot distance on the entanglement  of two electrons trapped in (In,Ga)As/GaAs QD molecules has been studied,\cite{He:2007} as well as the effect of ionization on the entanglement of two electrons in a single QD.\cite{Ferron:2009} 

In the present paper, we investigate how the geometrical changes in the confinement potential of single, core-shell and double QD structures  influence the spatial entanglement\cite{Coe:2008} between two electrons trapped within the nanostructure. To this aim we will use the two-center power-exponential potential\cite{Kwasniowski:2008} which allows to change the confinement potential with continuity from one structure to the others. 

We will show that small variations in the confinement potential can induce large changes in the entanglement and present a potential quantum phase transition between states with minimum and maximum entanglement.
We will perform a detailed study of the role of the Coulomb interaction in determining the entanglement and discuss the features of the many-body wave function which characterize the various entanglement regimes. 
  
The paper is organized as follows: the theoretical model is presented in section II; in section III we present and analyze the results for the system ground state energy and related Coulomb interaction; section IV includes the discussion of the entanglement; in section V we analyze the characteristics of the many-body wave function; in section VI we present entanglement indicators; and in section VII we discuss a potential quantum phase transition. Finally section VII  is devoted to conclusions and summary.
  
\section{Theoretical Model}
We consider two interacting electrons confined within a nanostructure and solve the related one-dimensional problem: we expect the  characteristics displayed by the entanglement within the single-dot and core-shell single-dot systems  to remain valid (at least qualitatively) for the corresponding spherically symmetric dots.
The entanglement related to a two-wells one-dimensional potential corresponds instead to the one generated within two separate spherically symmetric  quantum dots. This system in fact respects the topology of our one-dimensional model. As we will see, the analysis of the relationship between the many-body wave-function characteristics and the features of the entanglement supports this choice.
      
The Hamiltonian describing our  system is
\begin{equation}
	H=\sum_{i=1}^{2}{\left[-\frac{1}{2}\frac{d^{2}}{dx_{i}^{2}}+V(x_{i})\right]}+U(x_{1},x_{2}),
\end{equation}
where we have used (effective) atomic  units  and $U(x_{1},x_{2})=\delta(x_{1}-x_{2})$ models the Coulomb repulsion between two electrons in one dimension.\cite{delta_1D} 
The nanostructure  is modeled using the two-center power-exponential  potential\cite{Kwasniowski:2008} $V(x)$  given by
\begin{equation}
V(x)=-V_{0}\left\{\exp[-(|x+d|/R)^{p}]+\exp[-(|x-d|/R)^{p}]\right\}
\label{eqn:2centrePE} 
\end{equation}
with  the two centers symmetric in respect to the origin and situated at $\pm d$. This allows us to study a broad class of confinement potentials with different shape, size, softness and smoothness of the nanostructure boundaries.

By varying  its parameters, this potential can describe single and double  quantum dot structures---to model for example gate-defined or self-assembled QDs---as well as core-shell quantum dots (such as the ones synthesized using colloidal assembling techniques\cite{Dabbousi:1997}) or self assembled QDs with compositional modulation (see Table \ref{tbl:dots}).

\begin{table}
\centering
\begin{tabular}{ || c | c | c|| }
\hline
  Dot type & $R$ and $p$ range & Potential  \\ 
    & & (harder for larger $p$) \\ \hline \hline
Single dot\cite {Singledotgated, Singledotcolloidal, Singledotselfassembled} & $12\lesssim R$; $p\leq 2$ & Single well  \\ 
& $20\lesssim R$; $p\lesssim 4$  &  \\ \hline
Core-shell\cite{Dabbousi:1997,SecondColloidalcoreshell}& $9\lesssim R\lesssim 16$; $p > 2$  & Well within a well\\ 
 & $16\lesssim R\lesssim 30$; $p\gtrsim 7$ &\\ \hline
Double dot \cite{double_gate_dots,Doubledot2,Doubledot3}&   $R\lesssim 7$; $p\geq 1$ & Double well  \\ \hline
\hline
\end{tabular}
\caption{Table showing the main $R$ and $p$ parameter ranges corresponding to different types of nanostructures and potentials.}\label{tbl:dots}
\end{table}

\begin{figure}
	\centering
		\includegraphics[width=0.4\textwidth]{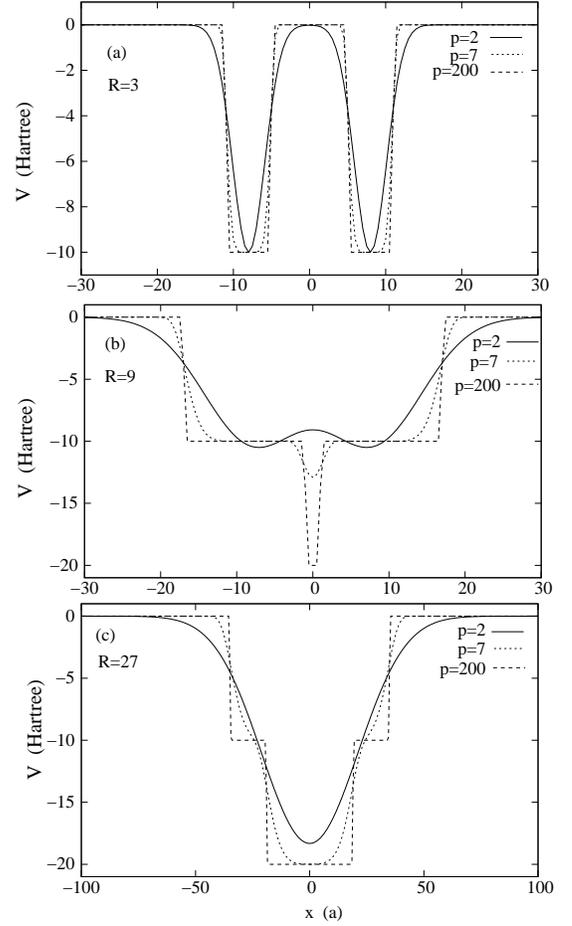}
	\caption{Confinement potential profile $V(x)$  as a function of $x$ and for different values of $R$. Solid lines correspond to  the soft confinement potential characterized by $p = 2$,  dashed lines correspond to  a rectangular-like confinement potential $(p = 200)$ and dotted lines correspond to an intermediate confinement potential $(p =7)$. Other parameters are $V_{0}=10$ Hartree and $d=8~a$.}
	\label{fig:Diagram2}
\end{figure}

Fig.~\ref{fig:Diagram2} highlights how the potential varies with the value of the parameters, and how its flexibility allows to consider `soft' or `hard' potentials as well as core-shells with different proportions between the two material components (compare for example center and bottom panel for $p=200$).

For fixed potential well depth $V_{0}$ and potential range $R$, the parameter $p$ in Eq.~(\ref{eqn:2centrePE}) characterizes the softness (hardness) of the confinement potential.  If  $p= 1$ the potential is triangular-like; for  $2\leq p \lesssim 10$ the confinement evolves from a `soft' Gaussian-like potential toward wells with a flat bottom and steep sides (`hard' potential). By increasing $p$ even further ($p \approx 200$) we obtain a potential practically indistinguishable from rectangular-like wells. 

Each panel in Fig.~\ref{fig:Diagram2} corresponds to a  different potential range $R$, whose value determines the transition between different geometries. For each $R$ we plot the potential for three different values of $p$ ($p=2$, $p=7$, and $p=200$). 

In the case of soft confinement potential ($p=2$), with increasing $R$ the resulting two-center potential changes smoothly from two separated wells to a single potential well. The smoothness of this potential can simulate for example the controlled, incremental doping of a semiconductor structure.  For $p=7$ we obtain a steeper confinement potential, but, more importantly, we obtain a `core-shell' structure for intermediate values of $R$. Notice that the fact that the potential is still relatively smooth can simulate the experimental situation in which there is a strong intermixing between the core and shell materials at their interface. For $p=200$  the confinement potential
is  rectangular-like and the structure intermediate between double and single well may correspond to colloidal  core-shell QD-nanostructures. In this case, we deal with a compound QD nanostructure, which consists of the small inner QD embedded in a larger outer shell with no relevant intermixing between the different materials.\cite{Dabbousi:1997}

To analyze  the influence of the geometry and softness of the confinement potential on the entanglement between two electrons trapped within the nanostructure, we  solve  the Schr\"odinger equation
\begin{equation}
H\Psi_i(x_{1},x_{2}) = E_i\Psi_i(x_{1},x_{2})	
\label{eqn:Schro}
\end{equation}
numerically by `exact' diagonalization.
To this purpose, we express the 
 wave function as a linear combination of single particle basis functions 
\be
\Psi_i(x_{1},x_{2})=\sum_{j_{1}}\sum_{j_{2}}a_{j_{1},j_{2};i}n_{j_{1}}(x_{1};\omega)n_{j_{2}}(x_{2};\omega),\label{eqn:Wavexpansion}	
\ee
where $\{n_{j_{i}}(x;\omega)\}$ are the eigenfunctions of the one-dimensional harmonic oscillator with angular frequency $\omega$.  To solve Eq.~(\ref{eqn:Schro}) we truncate the expansion in Eq.~(\ref{eqn:Wavexpansion}) by considering only terms with $1\le j_l\le N$, $l=1,2$, where $N$ is the single-particle basis size to be used in the calculation. 

 We have performed our calculations for fixed $V_{0} = 10$ (effective) Hartree and $d = 8a$, where $a$ is the (effective) Bohr radius. We are interested in the system ground state and we find that, using $\hbar\omega = 0.25$ (effective) Hartree, a single-particle basis size of $N = 50$ is large enough to achieve convergence.\cite{8.35}

For simplicity in the following we will refer to the system ground state and to its eigenvalue as 
$\Psi(x_{1},x_{2})$ and $E$ respectively.

\begin{figure}
\centering
	\includegraphics[width=0.4\textwidth]{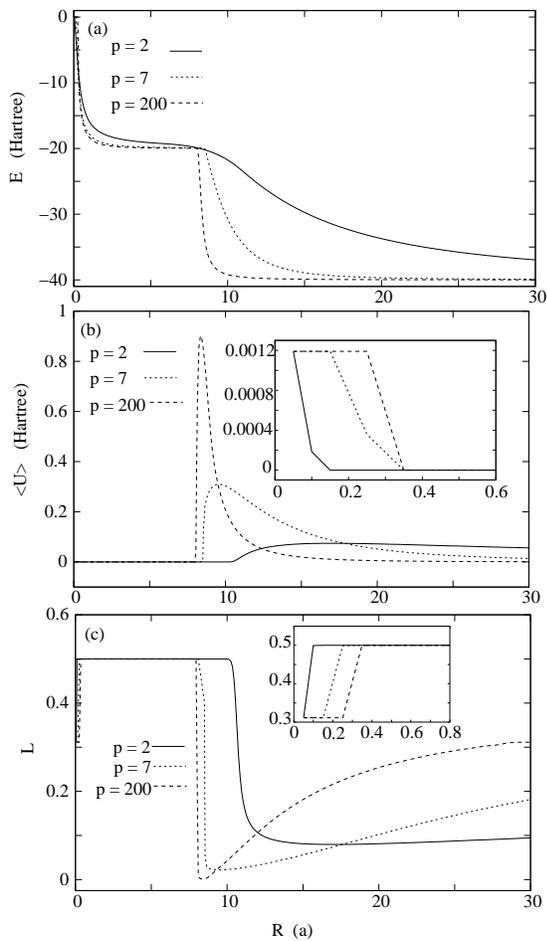}
	\caption{Ground state energy $E$ (upper panel), Coulomb repulsion $<U>$ (medium panel), and linear entropy of the reduced density matrix  $L$ (lower panel) versus confinement potential range $R$, for different values of $p$ (as labeled). The results for  $<U>$ and $L$ corresponding to small values of $R$ are shown in the insets of the middle and lower panel, respectively. }
	\label{fig:Diagram5}
\end{figure}

\section{Ground state energy and Coulomb repulsion}
To understand the entanglement properties of these systems, it will be helpful to consider first the behavior of the ground state energy $E$ and of its Coulomb repulsion component $<U>$. 

Figures \ref{fig:Diagram5}(a) and \ref{fig:Diagram5}(b) show the results for $E$ and $<U>$, respectively, as a function of the confinement potential range $R$ and different values of $p$.
For $p=2$ (`soft' confinement potential) and large $R$ the structure is a single well of depth $2V_{0}\exp(-d^2/R^2)$. The two electrons are localized at the bottom of the well with an energy $E\approx -4V_0\exp(-d^2/R^2)$. With decreasing $R$ the energy increases as the well narrows and its depth diminishes. For $R\approx 10$ the structure starts to  split into two wells of depth of approximately $V_{0}$, centered at $\pm d$ and  separated by a shallow potential barrier. The ground state energy is now $E\approx -2V_0$. The formation of the barrier between the wells produces a rapid decrease of the Coulomb interaction down to a negligible value. The height of the  barrier increases with decreasing  $R$, while the width of each well decreases maintaining a depth of $V_{0}$. This reflects in a slight increase of $E$.   For $R\approx 1$ the two wells become so narrow that there is  a rapid increase of the system kinetic energy and correspondingly a sharp variation in the ground state energy derivative: $E$ now rapidly increases to reach its maximum value. As $R$ decreases  the electronic wave-function starts to spread out of the wells. For  $R\lesssim 0.1$ this determines a slight increase of the Coulomb interaction (see inset in Fig.~\ref{fig:Diagram5}(b)).

For $p=7$ and $R\approx 30$ the confinement potential passes from a single well to a  core-shell QD nanostructure with a thin outer shell and a smooth transition between core and shell. As $R$ decreases, the external layer becomes thicker and more sharply defined, while the inner core diameter decreases in size.  The two electrons are localized at the bottom of the inner well with a corresponding ground state energy of $E\approx -4V_0$. The Coulomb repulsion increases with decreasing inner core width. 
For $9\lesssim R \lesssim 15$ the width as well as the depth of the inner core decreases and correspondingly both $E$ and $<U>$ rapidly increase. At $R\approx 10$ the electronic wave function start to spread into the outer shell and for $R<9$ the structure turns into a double well with  a depth of $V_0$ and an interwell barrier whose height increases with decreasing $R$. As the  electrons localize in different wells, the Coulomb interaction suddenly drops to a negligible value.   
For $R \lesssim 1$ the two separated wells become so narrow that  the system kinetic energy rapidly increases, similarly to $p=2$. For  $R\lesssim 0.3$ the spreading of  the electronic wave-function  outside the wells determines a slight increase of the Coulomb interaction (see inset in  Fig.~\ref{fig:Diagram5}(b)).

In the case of a rectangular-like confinement potential $(p=200)$, the type of nanostructures encountered for different $R$ are similar to the $p=7$ case. However the increased steepness of the potential walls and the abrupt transition between core and outer shell  and between the nanostructure and the surrounding material induce sharper transitions for decreasing $R$ in both the ground state energy and the Coulomb repulsion. The energy is basically  constant for $R \gtrsim 10$ and $2\lesssim R \lesssim 8$. For $R \approx 9$ the Coulomb interaction rapidly reaches its maximum as the inner core becomes very narrow but  the electrons are still localized within it. Then, as the electrons delocalize into the outer shell, the Coulomb repulsion suddenly decreases to a negligible value. It will slightly increase again for   $R\lesssim 0.3$ as the electronic wavefunction significantly spreads out of the wells.

We notice that, due to the chosen parameters, the Coulomb repulsion represents at any time a very small fraction of the total energy. However we will show that it will play a crucial role for the understanding of the behavior  of the entanglement.

\section{Entanglement and Linear Entropy}            
Bipartite entanglement is nowadays well-stated for distinguishable two-component quantum systems. In particular the state representing an entangled system cannot be factorized into a product of independent states describing its parts.  

Difficulties appear in classifying and quantifying the entanglement of a  system composed by indistinguishable particles. This is due to the  requirement for the antisymmetrization or symmetrization under particle exchange of the wavefunction  describing indistinguishable fermions or bosons, respectively. 
In \refon{Schliemann:2001,Ghirardi:2004} it has been shown that the unavoidable correlations due to the particle-exchange symmetry 
can be related (for fermions) to the Slater rank of the state. This corresponds to the minimum number of Slater determinants in which the state can be expanded.   For two  indistinguishable fermions the minimum possible Slater rank is one, and the related entanglement corresponds to the unavoidable antisymmetrization of the wave-function. This entanglement cannot be  used as a resource for quantum-information processing. 

In the system of two indistinguishable fermions we are considering, the entanglement is distributed over the spin and spatial degrees of freedom. In this paper we are interested in the entanglement of the ground state: the related many-body wave function  can be factorized into spin and spatial components and the entanglement of the two parts treated separately. 

In particular the spin component of the ground state is a singlet state and thus always maximally entangled: if we consider then the wave function corresponding to the minimum entanglement content for this system, $\phi(x_{1})\phi(x_{2})(|\uparrow_1\downarrow_2>-|\downarrow_1\uparrow_2>)$,  we see that the (constant) entanglement embedded in the spin degrees of freedom corresponds indeed to that minimum entanglement stemming from the antisymmetry requirements. Hence, as the spin entanglement is constant, in the following we will focus on the entanglement generated by the spatial degrees of freedom only.\cite{Coe:2008} 

We wish to study how the spatial entanglement changes as the geometry of the confinement potential is modified. 
We will quantify the spatial entanglement by using the linear entropy L of the one-particle reduced density matrix, which is a useful entanglement measure for a two-fermion system with a very large number of degrees of freedom:\cite{Buscemi:2007,Coe:2008} L has been shown to have a behavior very similar to the Von Neumann entropy\cite{VonNew} when quantifying  the particle-particle entanglement;\cite{Buscemi:2007,Coe:2008} at the same time L is much easier to calculate, especially for a system with a very large number of degrees of freedom.  

The linear entropy of the reduced density matrix is given by
\begin{equation}
\nonumber
L=Tr (\rho_{\text{red}}- \rho_{\text{red}}^2)=1-Tr \rho_{\text{red}}^2.
\end{equation}
In the continuous case the  reduced density matrix is given by
\begin{equation}
\nonumber
\rho_{\text{red}}(x_{1},x_{2})=\int \Psi^{*}(x_{1},x_{3})\Psi(x_{2},x_{3})dx_{3},
\end{equation}
with
\begin{equation}
\nonumber
\rho^{2}_{\text{red}}(x_{1},x_{2})=\int \rho_{\text{red}}(x_{1},x_{3})\rho_{\text{red}}(x_{3},x_{2})dx_{3}
\end{equation}
and
\begin{equation}
\nonumber
Tr\rho^{2}_{\text{red}}=\int \rho^{2}_{\text{red}}(x,x)dx. 
\end{equation}
The linear entropy  measures the entanglement in a pure state by giving an indication of the number and spread of terms in the Schmidt decomposition of the state. 

The numerical results for the entanglement  in respect to $R$ are displayed in Fig.~\ref{fig:Diagram5}(c). 
$L$ presents some general characteristics: a flat region with $L=0.5$ for small values of $R$, followed by a sharp drop to its minimum value at intermediate $R$, and by a partial recovery of the entanglement as $R$ increases further.  

Let us consider in more detail the case of a `soft' confinement potential $(p=2)$. For  $R \gtrsim 18$ the entanglement is lower, while the Coulomb interaction is higher, than for  $p=7$ and $p=200$.  This is due to the shape of the  $p=2$ potential (Fig.~\ref{fig:Diagram2}(c)) which is narrower toward the bottom of the well, where the electrons are confined. For the range $10 \lesssim R\lesssim 12$ as the single QD splits into two separate potential wells the entanglement increases to its highest value and the  Coulomb interaction quickly diminishes.  For smaller $R$ the overall structure corresponds to two separated QDs and the entanglement does not change. The Coulomb interaction is negligible in this range. For  $R\lesssim 0.1$ as the Coulomb interaction slightly increases, the entanglement decreases up to $L\approx 0.3$.

For $p=7$ and  large $R$ both entanglement and Coulomb interaction are intermediate between the cases of $p = 2$ and $p = 200$, $L$ decreasing and $<U>$ increasing with decreasing $R$.  At $R\approx 17.5$ the curves describing $<U>$ for $p=2$ and $p=7$ cross and so do the curves describing the  entanglement: from Figs.~\ref{fig:Diagram5}(b) and \ref{fig:Diagram5}(c) we note that all the crossings between the curves describing the Coulomb repulsion correspond to crossings between the entanglement entropy curves, showing the strong correlation between the behavior of the spatial entanglement and the Coulomb interaction. For smaller $R$,  as long as the inter-particle interaction increases, the entanglement decreases and drops to its minimum value, corresponding to the maximum of $<U>$. For $8\lesssim R \lesssim 9$, as the Coulomb repulsion drops to a negligible value,  the entanglement springs to its maximum value $L=0.5$, the same for all values of p.\cite{max_L}  $L$ remains constant for $0.3 \lesssim R \lesssim 8$, but for $R\lesssim0.3$, each of the separated wells becomes so narrow that the electronic wave function spreads out of the wells, so that the entanglement decreases again toward a value of $L=0.3$. An increase of the Coulomb interaction up to $<U>=12\times 10^{-4}$ Hartree accompanies this last variation of the entanglement. 

For $p = 200$ the entanglement behavior is similar to the previous case. However the rectangular-like confinement potential allows for a larger modulation of the electronic wave-function at large values of $R$, so in this region the entanglement is higher (and the Coulomb repulsion lower) than for $p = 2$ and  $p = 7$ . For $R \approx 8$ the inner well becomes so narrow and the electrons are so confined  that the entanglement drops to approximately zero. 
{\it The minimum of the entanglement corresponds to the maximum Coulomb repulsion}.   As the Coulomb repulsion drops to zero, the entanglement increases sharply to its highest value, while the QD nanostructure (and the electronic wave-function) splits into separate wells. For $R\lesssim0.3$ the entanglement decreases again similarly to the case of $p = 7$. Notice that the entanglement reaches a very similar value for $R\approx 0$ and $R\approx 30$.

\section{The many-body wave function\label{mb-wf}}
In order to explain the behavior of the entanglement for the different strengths of the confinement potential and values of the confinement potential range $R$, we will discuss the properties of the system many-body wave function.

\begin{figure}
	\centering
	\includegraphics[width=0.5\textwidth]{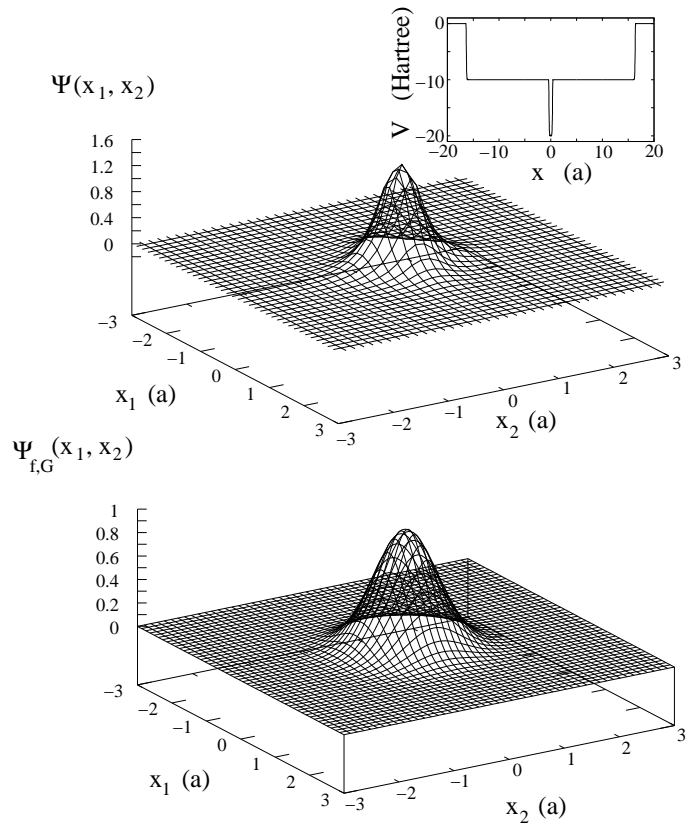}
	\caption{Upper panel: Two-electron wavefunctions $\Psi(x_1,x_2)$ for $R=8.35$ and $p = 200$. Inset: system confinement potential. Lower panel: factorized function $\Psi_{f,G}(x_1,x_2)=\exp(-2x_1^2)\exp(-2x_2^2)$}
	\label{fig:R=8.35}
\end{figure}

Let us consider first $p=200$ and the two extreme cases of maximum ($L=0.5$) and minimum ($L\approx 0$) entanglement.
The wave function corresponding to $L\approx 0$ ($R=8.35$) is shown in  Fig.~\ref{fig:R=8.35}, upper panel. Here the electrons are  confined within the very narrow core (see the upper panel inset, which shows the shape of the potential).
Due to this {\it very} strong-confinement regime, {\it even though the Coulomb repulsion is maximum}, the spatial part of the wave-function approaches the `non-interacting' uncorrelated limit $\Psi_f(x_1,x_2)=\phi(x_{1})\phi(x_{2})$, a product state which corresponds to no spatial entanglement. This is clearly shown by the comparison between the upper panel and the lower panel, which shows the Gaussian factorized state $\Psi_{f,G}(x_1,x_2)=\exp(-2x_1^2)\exp(-2x_2^2)$.

\begin{figure}
	\centering
	\includegraphics[width=0.4\textwidth]{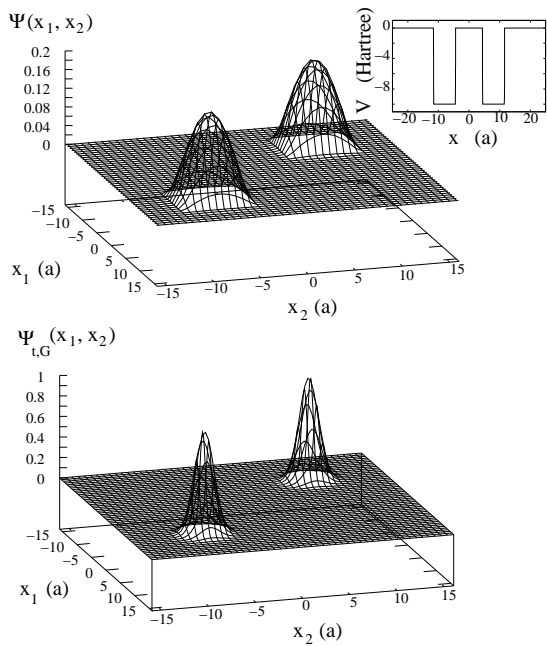}
	\caption{Upper panel: Two-electron wavefunctions $\Psi(x_1,x_2)$ for $R=3.6$ and $p = 200$.  Inset: system confinement potential. Lower panel: `triplet-type' function $\Psi_{t,G}(x_1,x_2)=\exp(-(x_1-8)^2/2)\exp(-(x_2+8)^2/2)+\exp(-(x_2-8)^2/2)\exp(-(x_1+8)^2/2)$.}
	\label{fig:R=3.6}
\end{figure}

At the opposite side of the entanglement spectrum, we find the system characterized by $R=3.6$ which corresponds to the maximum entanglement value $L=0.5$. In this case the nanostructure is composed by two relatively narrow and well-separate wells: each well would strongly confine the particles, but having more than one well provides an additional degree of freedom to the system. In this case even the weakest Coulomb interaction will then be able to spatially correlate the particles in a state with the structure $\Psi_{t}(x_1,x_2)=\Phi(x_1-d)\Phi(x_2+d)+\Phi(x_1+d)\Phi(x_2-d)$. This is the spatial equivalent of a spin triplet, and as such contains the same entanglement ($L=0.5$). Fig.~\ref{fig:R=3.6} shows the actual many-body wave function for $R=3.6$ (upper panel) and, for comparison, the Gaussian `triplet-type'  state $\Psi_{t,G}(x_1,x_2)=\exp(-(x_1-8)^2/2)\exp(-(x_2+8)^2/2)+\exp(-(x_2-8)^2/2)\exp(-(x_1+8)^2/2)$ (lower panel).  

\begin{figure}
	\centering
	\includegraphics[width=0.5\textwidth]{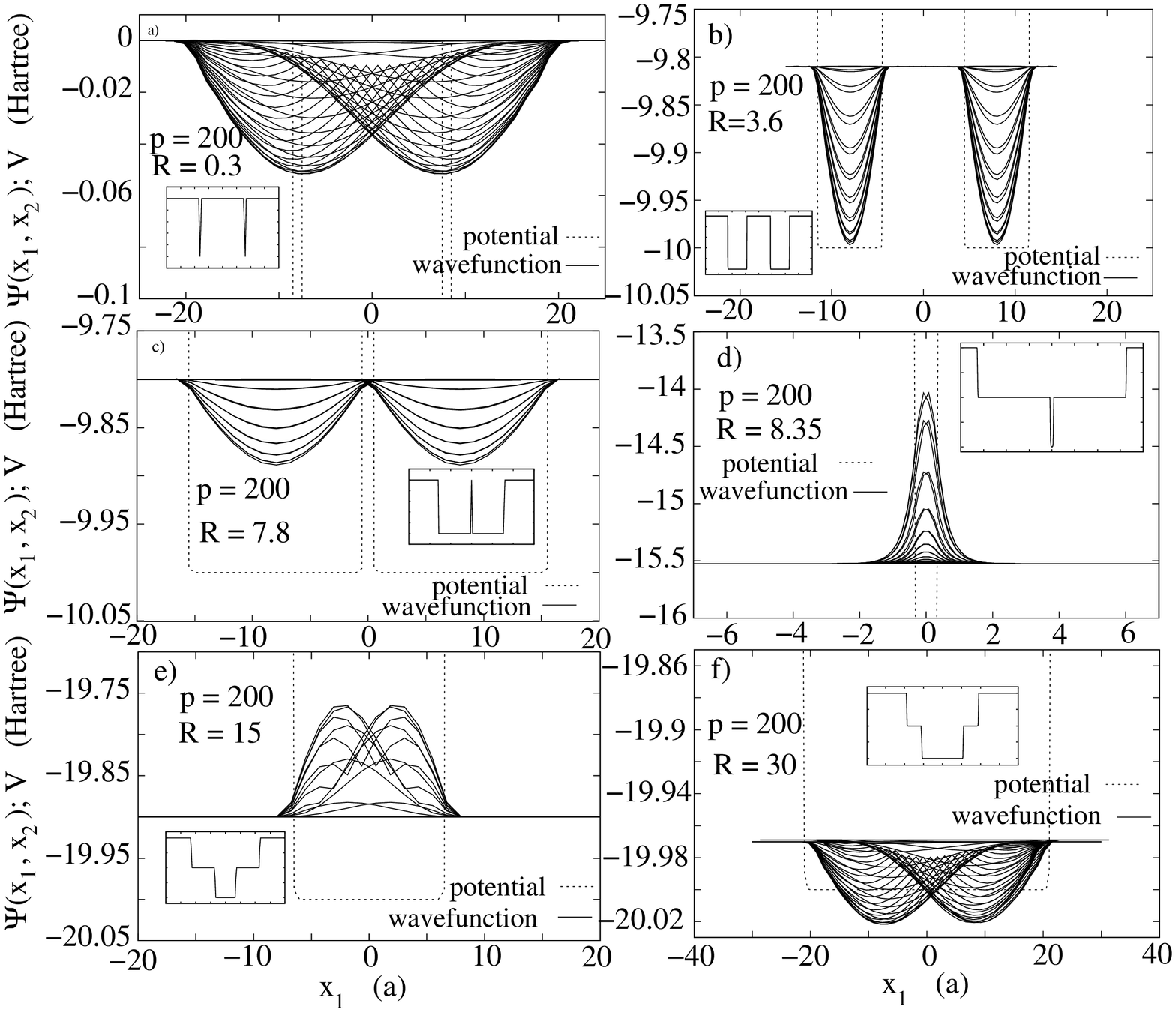}
	\caption{Two-electron wavefunctions over the confinement potential wells for six values of $R$ and  $p = 200$, as labeled. For each system $(R,p)$, the full confinement potential shape is shown in an inset. }
	\label{fig:Wavep=200}
\end{figure}

The other values of the entanglement correspond to wave-functions intermediate between the two described.
In Fig.~\ref{fig:Wavep=200} we present the wave-functions for $p=200$ and six values or $R$. Each wave function is a three-dimensional plot in the coordinates $(x_1,x_2,y=\psi(x_{1},x_{2}))$. We plot the projection of the wave functions on the plane $(x_1,y)$.
The wave function is drawn against the confining potential shape (dotted line) and, for each panel, the full confinement potential is presented in an inset.

The values of $R$ have been taken as follows: $R=0.3$ (Fig.~\ref{fig:Wavep=200}(a)) corresponds to the range where the two separated wells are extremely narrow. As a consequence the electronic wave function spills outside and spreads over a range much larger than the well boundaries. For this value of $R$ and  within the numerical limits of our calculation, electrons  are close to ionization ($E\approx 0$ Hartree).  The spreading of the wavefunction affects the entanglement properties of the system: the electrons are less localized, so that their spatial entanglement decreases while their Coulomb interaction increases. $R=3.6$ (Fig.~\ref{fig:Wavep=200}(b)) has been discussed before: the whole wave-function is localized close to the bottom of the two wells in a `triplet-type' state. Spatial entanglement is now maximum as the measure of one particle in a well would imply certain knowledge that the other particle is in the other well. The description is similar for  $R=7.8$ (Fig.~\ref{fig:Wavep=200}(c)), which corresponds to a state close to the edge of the region with constant $L=0.5$. Though the wave function shape is quite different from the case $R=3.6$, its overall structure remains the same and so does its entanglement. In this respect it is interesting to notice that a wide variety of nanostructures (see e.g. Fig.~\ref{fig:Diagram2}(a)) or insets in Fig.~\ref{fig:Wavep=200}(b) and (c)) would give rise to the same maximum entanglement. Surprisingly in this case the entanglement loses its strong sensitivity to the wave function details (see e.g. discussion in \refon{Coe:2008}) and the local shape of the potential does not affect the entanglement: Fig.~\ref{fig:Diagram5}(c) shows that the same value of $L$ is obtained for $0.5\lesssim R\lesssim8$ and $2\lesssim p\lesssim200$. Maximum entanglement seems indeed to be a very robust feature and to depend only on the topological feature of the wave function to be separable into two not-interconnected regions of space.  

$R=8.35$ corresponds, as mentioned, to the minimum value $L=0.3\times10^{-2}$: both electrons are strongly localized in the (same) core well and the spatial part of the wave function becomes close to a factorized form.       
We notice that the minimum of $L$ is very sharp (see Fig.~\ref{fig:Diagram5}(c)): in contrast to maximum entanglement, its minimum value can be achieved for very specific parameters and local confining potential shape only. As soon as the particles are less strongly confined, the system responds to Coulomb repulsion by increasing spatial correlation, which increases this type of entanglement. This explains why the minimum value of $L$ for the softer potentials characterized by $p=2$ and $p=7$ is actually higher than for $p=200$.  

As $R$ increases and the core well widens, the Coulomb repulsion forces the electrons apart and correlates their position further: this is evident when comparing Fig.~\ref{fig:Wavep=200}(d) with  Fig.~\ref{fig:Wavep=200}(e) and (f). Due to the enhanced correlation  the entanglement increases. 

We notice that, though the confining potential is very different, the wave-function (and its entanglement content) is very similar for $R=0.3$ and $R=30$. Again this shows that a similar entanglement within two electrons can be engineered by using very different types of nanostructures.  

The behavior of the wave functions for the softer potentials determined by $p=2$ and $p=7$ is similar. For $p=7$ the lesser strength of the wave-function confinement translates into a less  pronounced minimum in the entanglement; for $p=2$ the transition between single and double well is even smoother, and the ground state wave-function does not change significantly for $ 15\lesssim R \lesssim 30$ (not shown).  

We underline that both maximum and minimum entanglement correspond to a system which is strongly localized and strongly confined. The main difference between the two extremes is that to achieve the maximum entanglement some degrees of freedom which allow for correlation are necessary (more than one well in this case). 

\section{Entanglement indicators}  
\subsection{Probability density at the origin}
We will now show that the entanglement behavior can be inferred by considering the wavefunction behavior along two specific directions, $x_1=x_2$ and $x_1=-x_2$. 

\begin{figure}
	\centering
	\includegraphics[width=0.5\textwidth]{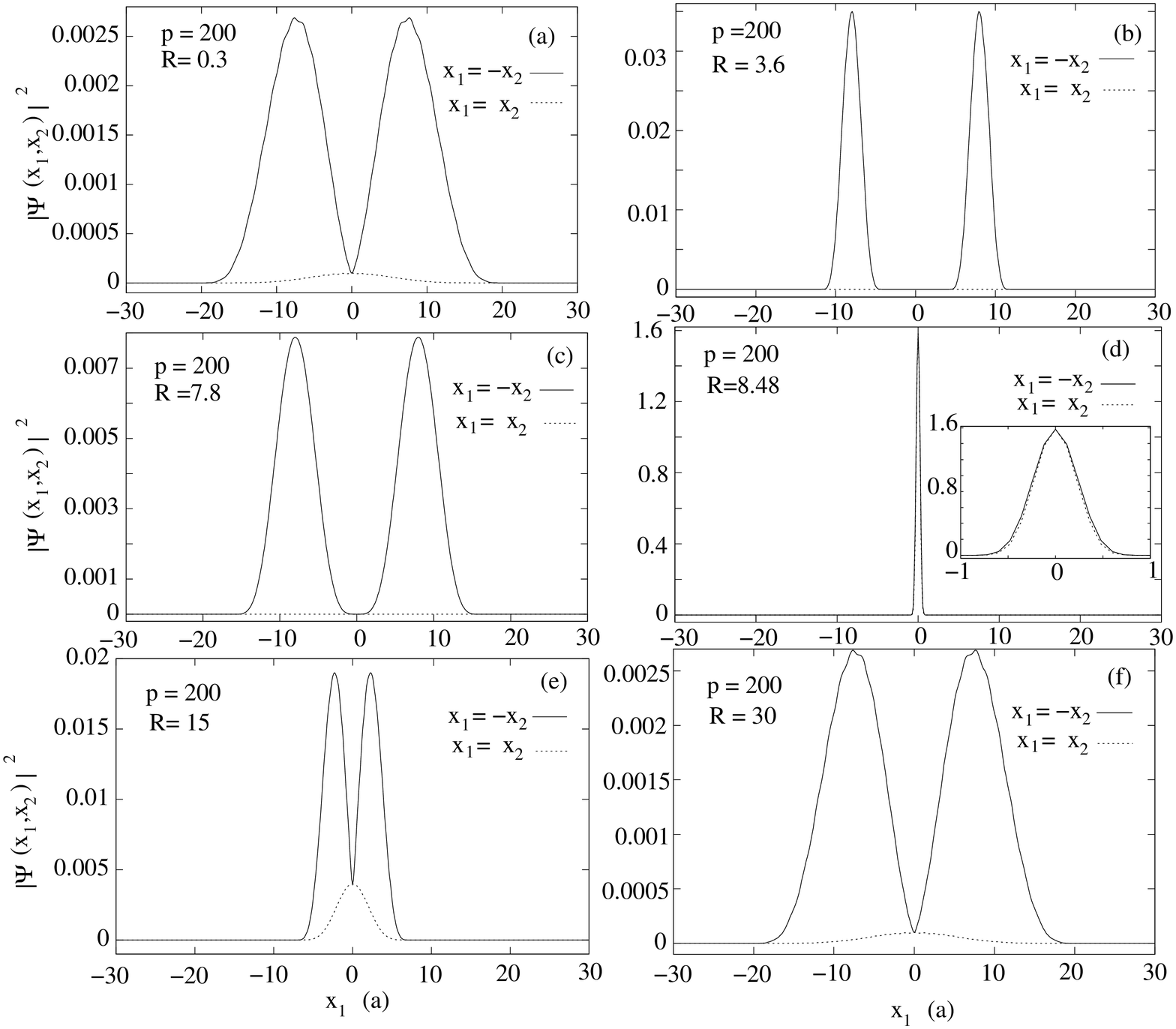}
	\caption{$|\Psi(x_1, x_2=x_1|^2)$ (dotted line) and $|\Psi(x_1, x_2=-x_1)|^2$ (solid line) in respect to $x_1$ for $p=200$ and the same sequence of values of $R$ as in Fig.~\ref{fig:Wavep=200}. For clarity a zoom of the wave-function cuts around $x_1=0$ are plotted in the inset of panel (d).  }
	\label{12}
\end{figure}

$|\Psi(x_1, x_2=x_1)|^2$ represents the probability density of
finding both electrons at the same point, while  $|\Psi(x_1, x_2=-x_1)|^2$ represents the probability density of
finding the electrons at two different points which are symmetric in respect to the $y$-axis. 
In Fig.~\ref{12} we plot $|\Psi(x_1, x_2=x_1)|^2$ (dotted line) and $|\Psi(x_1, x_2=-x_1)|^2$ (solid line) for $p=200$ and the same sequence of $R$ values as in Fig.~\ref{fig:Wavep=200}

Low spatial entanglement implies that there is a high probability of finding both particles at the same position $x$: this would in fact imply that Coulomb repulsion is unable to induce spatial correlation and the system is close to a factorized state. Vice-versa we expect
 $|\Psi(x_1, x_2=x_1)|^2=0$  (and, in particular, $|\Psi(0, 0 )|^2=0$) for a maximally entangled state, as confirmed by Fig.~\ref{12}(b) and (c).  

High spatial entanglement will imply that finding one particle at $x$ will inform us that the other particle has a high probability to be at a different, but specific, location, i.e., by the symmetry of our confinement potential, at $-x$.  We expect then $|\Psi(x_1, x_2=-x_1)|^2$ to present peaks symmetric in respect to $x_1=0$ when the spatial entanglement is non-zero, as shown in Fig.~\ref{12}.

For $R$ corresponding to minimum entanglement,  $|\Psi(x_1, x_2=-x_1)|^2\approx|\Psi(x_1, x_2=x_1)|^2 $ and $|\Psi(0,0)|^2$ reaches its maximum (Fig.~\ref{12}(d)). The lack of differentiation between these two `orthogonal' type of correlations (having the same probability of finding a particle at the same position or in a position symmetric in respect to the origin) indicates in fact that little information on the other particle can be gained by measuring the position of one of the particles. 
By looking at Fig.~\ref{12}(d) we notice  once more that, for minimum entanglement the wave function assumes a form close to the factorized form $\Psi(x_1, x_2=x_1)\approx\Phi(x_1)\Phi(x_2)$.
 
\begin{figure}
	\centering
	\includegraphics[width=0.5\textwidth]{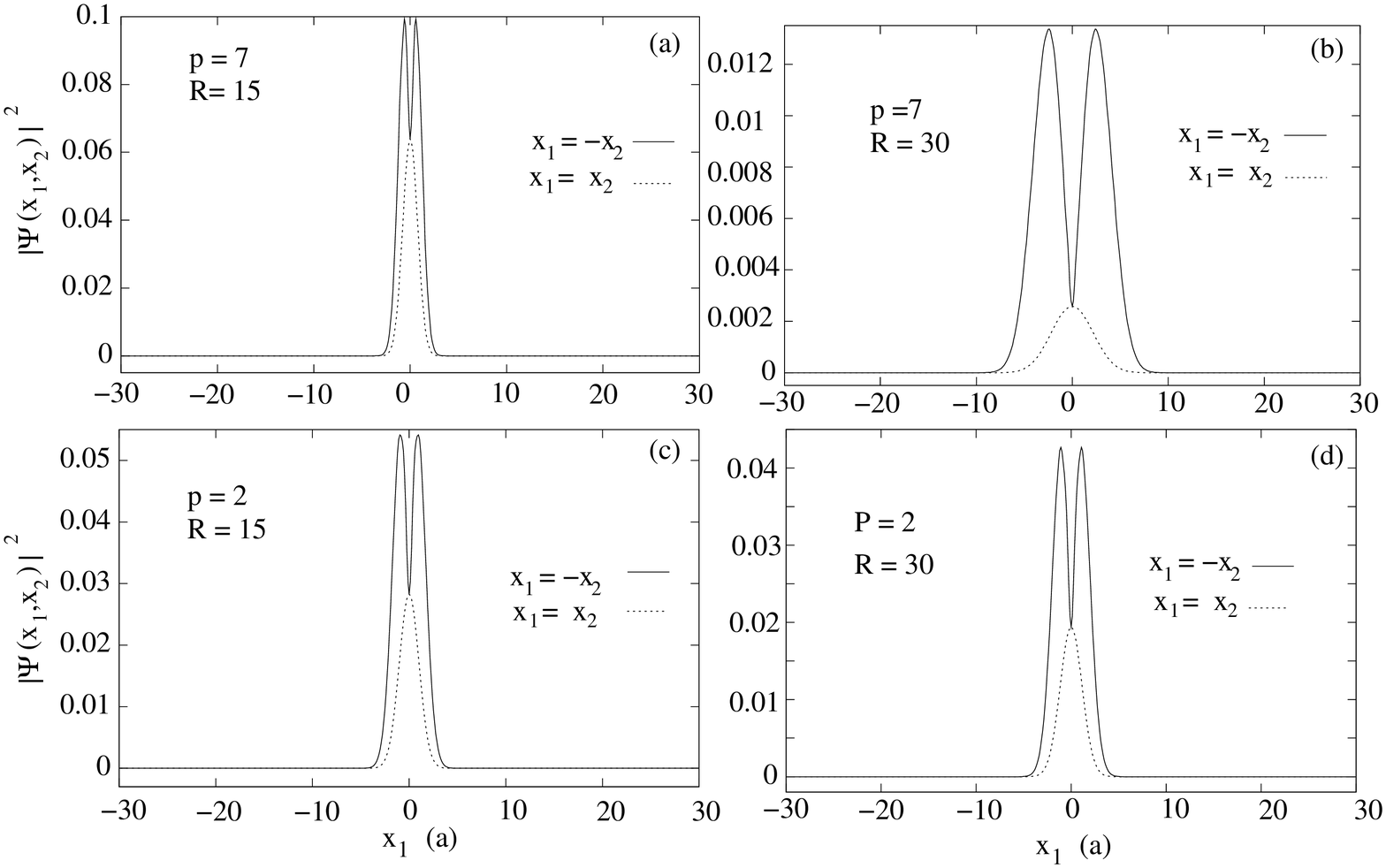}
	\caption{$|\Psi(x_1, x_2=x_1|^2)$ (dotted line) and $|\Psi(x_1, x_2=-x_1)|^2$ (solid line) in respect to $x_1$ for $p=2$ and $p=7$,  and $R=15$ and $R=30$ }
	\label{13}
\end{figure}

In general the value of the probability density $|\Psi(x_1, x_2)|^2$ at the highly symmetric point $x_1= x_2=0$ (common to both the directions considered) will increase for decreasing entanglement: let us consider the values of $L$ and $|\Psi(0,0)|^2$ for $R=15$ and the different values of $p$. From Fig.~\ref{fig:Diagram5}(c) we see that $L(p=7,R=15)<L(p=2,R=15)<L(p=200,R=15)$. By comparing Fig.~\ref{13}(a), (c) and Fig.~\ref{12}(e) we find that instead $|\Psi(0,0;p=7,R=15)|^2>|\Psi(0,0;p=2,R=15)|^2>|\Psi(0,0;p=200,R=15)|^2$. If we now consider $R=30$, the $p=2$ and $p=7$ curves for the entanglement have crossed, and indeed we find that  $L(p=2,R=30)<L(p=7,R=30)<L(p=200,R=30)$, and, from Fig.~\ref{13}(b), (d) and Fig.~\ref{12}(f),  $|\Psi(0,0;p=2,R=30)|^2>|\Psi(0,0;p=7,R=30)|^2>|\Psi(0,0;p=200,R=30)|^2$.
These findings confirm that the value of the probability density at a single but highly symmetric point contains very relevant information on the value of the overall system entanglement.  False indication of entanglement might occur when considering mixed states, but similar problems would be encountered when using pure-state entanglement measures such as L or the Von Neumann entropy, and even criteria designed for mixed states will not always detect bipartite entanglement between systems with large numbers of degrees of freedom.\cite{Bruss:2002}

\subsection{Coulomb interaction and effects of long-range repulsion}
So far we have modeled the Coulomb repulsion as a contact interaction. This implies that  the Coulomb repulsion is non zero only if the wave function allows for both particles to be at the same place. The higher the probability of having particles at the same position, the higher the Coulomb repulsion and the lower the entanglement. 
This counterintuitive `inverse' relationship between  Coulomb repulsion and entanglement
behaves as a very good entanglement indicator, as can be observed by comparing Fig.~\ref{fig:Diagram5}(b) and (c), where the maxima (minima) of the Coulomb interaction corresponds to the minima (maxima) of the entanglement. We may wonder though if this is an artefact of the contact-type of interaction used.

\begin{figure} 
		\includegraphics[width=0.4\textwidth]{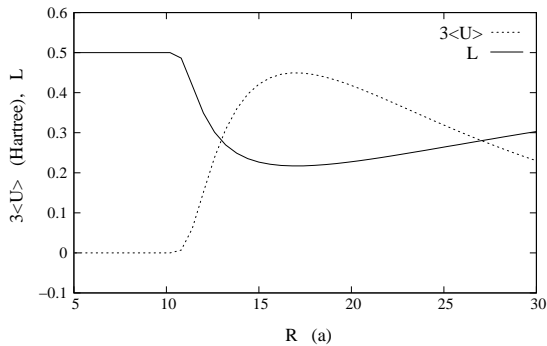}
	\caption{Coulomb repulsion $<U>$ and entanglement entropy  $L$ versus potential range $R$ for the long-range Coulomb repulsion \ref{longrangeC} and the Gaussian-type confinement potential $p = 2$}
	\label{fig:longrangeC}
\end{figure}

In Fig.~\ref{fig:longrangeC} we present the calculations done using  the long range Coulomb repulsion
\be
U_{lr}(x_1,x_2) = {1\over\sqrt{1+(x_1-x_2)^2}}\label{longrangeC}
\ee
and a basis size $N=30$, which allows for convergency for the range of values of $R$ shown. We choose $p=2$ which, describing the softest potential among the set of $p$-values chosen, would allow for the biggest modification of the corresponding many-body wave-functions, and hence of the entanglement.

Fig.~\ref{fig:longrangeC} shows first of all that the main characteristics of the entanglement found with the contact interaction are confirmed: the entanglement entropy presents a plateau with $L=0.5$ for $R\lesssim10$, a rapid decrease for intermediate values of $R$ reaching a minimum for $R\approx 17$, and a slow increase of the entanglement as $R$ is increased further.
The higher value of the minimum, in respect to the results obtained using the contact-type of interaction, witnesses the increase of spatial correlations due to the long range nature of Eq.~(\ref{longrangeC}). 
 
Most importantly though, the results obtained using a long range interaction still show the same `inverse' correlation between Coulomb repulsion and entanglement:
this indicates that Coulomb repulsion between particles, is a good indicator for spatial entanglement.

\section{Potential quantum phase transition}
A point of nonanalyticity in the ground state energy of a quantum system is associated with a quantum phase transition.\cite{QPT}  In \refon{Wu:2006} it was shown that in such a transition a nonanalyticity of the entanglement measure would be associated to the nonanalyticity of  the ground state energy. 

In the system we are considering, for increasing $p$, as the confining potential becomes harder and harder, a discontinuity in the derivative  in respect to the potential range of the ground state energy and of the entanglement measure,  $\partial E/\partial R$ and $\partial L/\partial R$, seems to appear. This discontinuity underlines the transition between minimum and maximum values of the entanglement (see $R\approx 8$ for $p=200$).

A similar pattern for the  entanglement was observed in the quantum phase transition for two electrons close to the ionization point of a single QD.\cite{Ferron:2009}  In that case the transition was between bound and unbound (resonance) states, while in our case a transition seems to occur between two different sets of bound states. 
 
In the system we are considering, the transition is triggered by a shape-change in the confining potential, from the potential in the inset of  Fig.~\ref{fig:Wavep=200}(d) to the one in the inset of  Fig.~\ref{fig:Wavep=200}(c). 
Due to this change, the system shifts between two very different  sets of ground states: the first set describes  the particles being highly confined in the narrow core region of a core-shell type structure, and it is formed by wave functions similar to the factorized $\Psi_f(x_1,x_2)$; the second set describes particles confined within two separate wells and it is formed by the topologically different `triplet-type' states $\Psi_{t}(x_1,x_2)$.  The fact that the system ground state on the left and right of the transition has well defined but very different properties is consistent with a quantum phase transition picture (see e.g. \refon{Gu:2004}).

We note that as $R$ decreases and the transition is approached, the energy difference between the system ground state, as bounded within the inner well, and the ground state that the system would have if the inner well would be absent (and which would correspond to a double well with a barrier of vanishing width\cite{possibleQPTfootnote}) decreases as well. This energy distance between the two relevant sets of bound states reaches a minimum at the transition point.

The transition between these two confinement potential shapes could be induced experimentally by changing the polarity of a gate positioned over the core-to-barrier region within a gate-defined QD.  Recent studies have shown the possibility of engineering gate-defined QD confinement potentials with shapes ranging from Gaussian, to rectangular-type potentials, \cite{PhysE:2003} so this type of device should allow the transition between minimum and maximum entanglement to be explored even in (or at least close to) the large $p$ limit.
\section{Conclusions}
We have studied the entanglement of two interacting electrons confined within single, core-shell and double quantum dots.  The confinement potential has been parametrized by a two-center power-exponential potential, which has allowed us to investigate quantum dots described by either hard or soft potentials, with different ranges, including the effects of the transition between the different types of structures.  The calculation has been done by direct diagonalization of the Hamiltonian  including the Coulomb interaction between the electrons.  By varying the confinement potential as a function of dot shape, range of confinement potential---which determines the QD size---and the strength of the confining potential we showed that it is possible to induce fast and large variations of the entanglement between the two electrons.  This property may be used to design nanostructures---and nanostructure modulations via external fields---according to the level of entanglement required by a specific application.  

We have studied in detail the relationship between Coulomb repulsion and spatial entanglement and shown that they display a counterintuitive `inverse' correlation: due to Coulomb repulsion, electrons tend to correlate their position to {\it minimize} their interaction, which implies minimizing the probability that electrons could be found at the same position. From the entanglement point of view this correlation means enhancing the probability that if one electron is measured at a certain position the other will be in a different but correlated region, thus enhancing the spatial entanglement between the two particles.  We note however that, if the Coulomb interaction is switched off, the spatial entanglement vanishes (see also \refon{Lambert:2007}). This is in contrast to the behavior of the `local' (or `site')  entanglement which characterizes the Hubbard model, where zero Coulomb interaction corresponds to maximum entanglement.\cite{Gu:2004,Franca2:2006}

We have analyzed the many-body wave function and in particular the correlations between the particle probability density along some specific directions and the entanglement. We have then proposed the value of the particle probability density at a single (but highly symmetric point) as an indicator of entanglement.       

We have identified a potential quantum phase transition between minimally and maximally  entangled states within our system. This transition is triggered by a change in the potential shape which induces a  topological change in the many-body wave function, from an almost factorized to a `triplet-type' wave-function form.  Further investigation of this intriguing phenomena will be pursued in future work. 

Systems of electrons confined in quantum dots have been proposed as tools for performing quantum information/computation tasks; it is then of great importance to understand how the entanglement between these particles can be engineered and tailored. Our work provide a systematic study in this direction. 
\section*{ACKNOWLEDGMENT}
We gratefully acknowledge partial support from EPSRC through Grant No. EP/F016719/1.


\begin{thebibliography}{99}
\bibitem{QDSchemes} See, for example, E. Biolatti, I. D'Amico, P. Zanardi, and F. Rossi, Phys. Rev. B {\bf 65}, 075306 (2002); T. Tanamoto, Phys. Rev. A {\bf 61}, 022305 (2000); S. De Rinaldis, I. D'Amico, E. Biolatti, R. Rinaldi, R. Cingolani, and F. Rossi, Phys. Rev. B  {\bf 65}, 081309(R) (2002); T. E. Hodgson et al, J. Appl. Phys. {\bf 101}, 114319 (2007); X. Q. Li et al, Science {\bf 301}, 809 (2003); T. P. Spiller, I. D'Amico, and B. W. Lovett, New J. Phys. {\bf 9}, 20 (2007); A. Kolli, B. W. Lovett, S. C. Benjamin, and T. M. Stace, Phys. Rev. Lett. {\bf 97}, 250504 (2006); Gang Chen, N. H. Bonadeo, D. G. Steel, D. Gammon, D. S. Katzer, D. Park, and L. J. Sham, Science {\bf 289}, 1906 (2000); A. Imamog\={}lu, D. D. Awschalom, G. Burkard, D. P. DiVincenzo, D. Loss, M. Sherwin, and A. Small, Phys. Rev. Lett. {\bf 83}, 4204 (1999)
\bibitem{LdV} D. Loss and D. P. DiVincenzo, Phys. Rev. A {\bf 57}, 120 (1998)
\bibitem{laser} See, for example, A. J. Ramsay, R. S. Kolodka, F. Bello, P. W. Fry, W. K. Ng, A. Tahraoui, H. Y. Liu, M. Hopkinson, D. M. Whittaker, A. M. Fox, and M. S. Skolnick, Phys. Rev. B {\bf 75}, 113302 (2007); A. J. Ramsay, S. J. Boyle, R. S. Kolodka, J. B. B. Oliveira, J. Skiba-Szymanska, H. Y. Liu, M. Hopkinson, A. M. Fox, and M. S. Skolnick, Physical Review Letters {\bf 100}, 197401 (2008); D. J. Reilly, J. M. Taylor, J. R. Petta, C. M. Marcus, M. P. Hanson, and A. C. Gossard, Science {\bf 321}, 817  (2008); M. Stopa, and C. M. Marcus, Nano Lett. {\bf 8}, 1778  (2008); C. Barthel, D. J. Reilly, C. M. Marcus, M. P. Hanson, and A. C. Gossard, Phys. Rev. Lett. {\bf 103} 160503 (2009).
\bibitem{tune}I. D'Amico and F. Rossi, Appl. Phys. Lett. {\bf 79}, 1676 (2001); S. De Rinaldis, I. D'Amico, and F. Rossi, Appl. Phys. Lett. {\bf 81}, 4236 (2002); Patrik Recher, Johan Nilsson, Guido Burkard, and Bj\"{o}rn Trauzettel, Phys. Rev. B {\bf 79}, 085407 (2009)
\bibitem{double_gate_dots}J. M. Elzerman, R. Hanson, J. S. Greidanus, L. H. Willems van Beveren, S. De Franceschi, L. M. K. Vandersypen, S. Tarucha, and L. P. Kouwenhoven, Phys. Rev. B  {\bf 67}, 161308(R) (2003).
\bibitem{experimental_papers} Z. M. Wang , K. Holmes , Y. I. Mazur , and G. J. Salamo , Appl. Phys. Lett. {\bf 84}, 1931 (2004); S. Kiravittaya, A. Rastelli, and O. G. Schmidt, Reports on Progress in Physics {\bf 72}, 046502 (2009); L. Wang, A. Rastelli, S. Kiravittaya, M. Benyoucef, and O. G. Schmidt, Advanced Materials {\bf 21}, 2601 (2009).
\bibitem {e-m_entangl} P. I. Tamborenea and H. Metiu, Europhys. Lett. {\bf 53}, 776 (2001);
J. Schliemann, D. Loss, and A. H. MacDonald, Phys. Rev. B {\bf 63}, 085311 (2001); Stefan Legel, J\"urgen K\"onig, Guido Burkard, and Gerd Sch\"on, Phys. Rev. B {\bf 76}, 085335 (2007).
\bibitem {He:2007} L. He and A. Zunger, Phys. Rev. B {\bf 75}, 075330 (2007).
\bibitem{Ferron:2009} A. Ferr\'on, O. Osenda and P. Serra, Phys. Rev. A {\bf 79}, 032509 (2009).
\bibitem {Coe:2008} J. P. Coe, A. Sudbery, and I. D'Amico, Phys. Rev. B {\bf 77}, 205122 (2008).
\bibitem {Kwasniowski:2008} A. Kwa\'{s}niowski and J. Adamowski, J. Phys. Condens. Matter {\bf 20}, 215208 (2008).

\bibitem {delta_1D} R. J. Magyar and K. Burke, Phys. Rev. A {\bf 70}, 032508 (2004);  M. Casula, D. M. Ceperley and E . J. Mueller, Phys. Rev. A {\bf 78}, 033607 (2008); R. J. Magyar, Phys. Rev. B {\bf 79}, 195127 (2009).


\bibitem{Dabbousi:1997}B. O. Dabbousi, J. Rodriguez-Viejo, F. V. Mikulec, J. R. Heine, H. Mattoussi, R. Ober, K. F. Jensen, and M. G. Bawendi, J. Phys. Chem. B {\bf 101}, 9463 (1997).
\bibitem{Singledotgated} R. C. Ashoori, H. L. Stormer, J. S. Weiner, L. N. Pfeiffer, S. J. Pearton, K. W. Baldwin, and K. W. West, Phys. Rev. Lett. {\bf 68}, 3088 (1992).
\bibitem{Singledotcolloidal} A. A. Lalayan, Appl. Surf. Science {\bf 248}, 209 (2005).
\bibitem{Singledotselfassembled} A. Rastelli, M. Stoffel, A. Malachias, T. Merdzhanova, G. Katsaros, K. Kern, T. H. Metzger and O. G. Schmidt, Nano Lett. {\bf 8}, 1404 (2008).
\bibitem{SecondColloidalcoreshell} O. Millo, D. Katz, Y. W. Cao and U. Banin, Phys. Rev. Lett. {\bf 86}, 5751 (2001).

\bibitem{Doubledot2}J. R. Petta, A. C. Johnson, J. M. Taylor, E. A. Laird, A. Yacoby, M. D. Lukin, C. M. Marcus, M. P. Hanson, and A. C. Gossard, Science {\bf 309}, 2180 (2005).
\bibitem{Doubledot3} P. D. Siverns, S. Malik, G. McPherson, D. Childs, C. Roberts, R. Murray, B. A. Joyce and H. Davock, Phys. Rev. B {\bf 58}, R10127 (1998).  


\bibitem {8.35}To achieve very accurate results for the detail study of the minimum of the entanglement ($p=200$ and $R=8.35$) we have used a basis corresponding to the angular frequency $\hbar\omega = 16$ (effective) Hartree.
\bibitem {Schliemann:2001} J.Schliemann, J. I. Cirac, M. Kus, M. Lewenstein, and D. Loss,  Phys. Rev. A  {\bf 64}, 022303 (2001). 
\bibitem {Ghirardi:2004} G. C. Ghirardi and  L. Marinatto, Phys. Rev. A {\bf 70}, 012109 (2004).

\bibitem {Buscemi:2007} F. Buscemi, P. Bordone, and A. Bertoni, Phys. Rev. A {\bf 75}, 032301 (2007).
\bibitem {VonNew}M. A. Nielsen and I. L. Chuang, `Quantum Computation and Quantum  Information', Cambridge University Press (2000)   
\bibitem {max_L}  The theoretical maximum value for the spatial entanglement is $L=1$, see \refon{Coe:2008}, but the maximum value reachable for the ground state of the type of nanostructures we are considering is $L=0.5$.


\bibitem{Bruss:2002} D. Bru\ss, J. Math Phys {\bf 43}, 4237 (2002).


\bibitem {QPT} S. Sachdev, `Quantum Phase Transitions', Cambridge University Press (2000).
\bibitem {Wu:2006}L. A. Wu, M. S. Sarandy, D. A. Lidar, and L. J. Sham, Phys. Rev. A {\bf 74}, 052335 (2006). 

\bibitem{Gu:2004} S.-J. Gu, S.-S Deng, Y.-Q. Li, and H.-Q. Lin, Phys. Rev. Lett. {\bf 93}, 086402 (2004).
\bibitem{possibleQPTfootnote}  In this respect we note that a related abrupt entanglement transition connecting two different sets of bound states, should occur even if the structure considered for $R$ smaller than the transition value would simply be a larger single well (the `outer' well), whose width for example increases for decreasing $R$. In this case we would expect though that the entanglement value corresponding to $R$ smaller than the transition point would not be constant and should be less than $0.5$, as the corresponding ground state wavefunctions would be intermediate between the `factorized' and `triplet-type' forms.


\bibitem{PhysE:2003} K. Lis, S. Bednarek, B. Szafran and J. Adamowski, Physica E: Low-dimensional Systems and Nanostructures {\bf 17}, 494 (2003). 



\bibitem{Lambert:2007} N. Lambert, R. Aguado, and T. Brandes, Phys. Rev. B {\bf 75}, 045340 (2007).



\bibitem{Franca2:2006} V. V. Fran\c{c}a and K. Capelle, Phys. Rev. A {\bf 74}, 042325 (2006).
\end{thebibliography}
\end{document}